\documentclass[secnumarabic,amssymb, nobibnotes, aps, prd]{revtex4-2}
\usepackage{tocloft}
\usepackage{graphicx}
\usepackage{amsmath}
\usepackage{subcaption}
\usepackage{hyperref}
\hypersetup{colorlinks,linkcolor={red},citecolor={blue},urlcolor={red}}
\usepackage{setspace}
\usepackage{xcolor}
\pagecolor{white}

\usepackage{multirow}
\color{black}

\usepackage{graphicx}
\usepackage{amsmath}
\usepackage{subcaption}
\usepackage{hyperref}
\hypersetup{colorlinks,linkcolor={red},citecolor={blue},urlcolor={red}}
\usepackage{setspace}
\usepackage{xcolor}
\pagecolor{white}

\usepackage{multirow}

\begin{document}
	\title{The Interconnection of Cosmological Constant and Renyi Entropy in Kalb-Ramond Black Holes : Insights from Thermodynamic Topology}
	
	\author{Bidyut Hazarika$^1$}
	\email{$rs_bidyuthazarika@dibru.ac.in$}
		
	\author{Mozib Bin Awal$^1$}
	
	\email{$mozibawal@gmail.com$}
	
	\author{Prabwal Phukon$^{1,2}$}
	\email{$prabwal@dibru.ac.in$}
	
	\affiliation{$^1$Department of Physics,Dibrugarh University, Dibrugarh,Assam,786004.\\$^2$Theoretical Physics Division, Centre for Atmospheric Studies, Dibrugarh University, Dibrugarh,Assam,786004.\\}
	\begin{abstract}
		This paper seeks to establish a connection between the cosmological constant and Renyi entropy within the framework of Kalb-Raymond(K-R) gravity. Our analysis is supported by evidence showing the equivalence of the thermodynamic topology of K-R AdS black holes in the Gibbs-Boltzmann (GB) statistical framework and K-R flat black holes in the Renyi statistical framework. We begin by exploring the thermodynamic topology of K-R black holes in flat spacetimes, focusing on the topological characteristics and phase transition behavior in both statistical frameworks. We find that K-R flat black holes in Renyi statistics exhibit equivalent global and local topological properties to K-R AdS black holes in GB statistics. This equivalence points to a potential connection between the cosmological constant and the Renyi parameter. We derive an approximate relationship between the Renyi parameter and the cosmological constant, which is consistent with similar findings in the literature from a cosmological perspective.
	\end{abstract}
	
	\maketitle
	
	
	
	
	
\section{introduction}

Black holes, among the most enigmatic phenomena in the universe, are regions of intense spacetime curvature defined by an event horizon-a boundary beyond which neither matter nor light can escape. Black hole thermodynamics has evolved significantly since the 1970s, with major milestones attributed to the groundbreaking work of Bekenstein and Hawking. Bekenstein’s concept of black hole entropy, combined with Hawking’s groundbreaking discovery of black hole radiation, established a deep connection between black holes and thermodynamic principles \cite{Bekenstein:1973ur,Hawking:1974rv,Hawking:1975vcx}. These discoveries transformed black holes from purely astrophysical phenomena into objects governed by thermodynamic laws, initiating a major shift in theoretical physics \cite{Bardeen:1973gs}. Following these foundational contributions, numerous studies have expanded on the intricate relationship between black holes and thermodynamics \cite{Wald:1979zz,bekenstein1980black,Wald:1999vt,Carlip:2014pma,Wall:2018ydq,Candelas:1977zz,Mahapatra:2011si}.  A central feature of black hole thermodynamics is the study of phase transitions. Davies was the first to identify phase transitions through discontinuities in heat capacity at specific points \cite{Davies:1989ey}. Another crucial development is the Hawking–Page transition, which occurs when the free energy of the black hole shifts, indicating a transition between distinct thermodynamic states \cite{Hawking:1982dh}. Furthermore, various scenarios involving transitions from non-extremal to extremal black hole states have been extensively examined \cite{curir_rotating_1981,Curir1981,Pavon:1988in,Pavon:1991kh,OKaburaki,Cai:1996df,Cai:1998ep,Wei:2009zzf,Bhattacharya:2019awq}.  In addition, analogies with van der Waals systems have provided valuable insights into black hole phase transitions and critical behavior, deepening our understanding of their thermodynamic properties \cite{Kastor:2009wy,Dolan:2010ha,Dolan:2011xt,Dolan:2011jm,Dolan:2012jh,Kubiznak:2012wp,Kubiznak:2016qmn,Bhattacharya:2017nru}. These parallels have further enriched the study of black hole thermodynamics, offering a broader perspective on critical phenomena and phase transitions.\\

In black hole physics, thermodynamic entropy is fundamentally linked to the surface area of the event horizon. If black holes are viewed as three-dimensional objects, this area-based scaling conflicts with the conventional notion of extensive thermodynamic entropy. As a result, the classical Boltzmann-Gibbs statistical framework may be insufficient for accurately describing black hole thermodynamics, indicating the necessity for alternative approaches. To gain a deeper understanding of black hole entropy, various extensions of Boltzmann-Gibbs statistics have been explored in the literature \cite{Cirto,Quevedo,Tsallis,Barrow,Nojiri,Reny,Sharma,Kania0}.
Over the past few decades, developments in quantum mechanics have led to modifications in the entropy associated with black hole horizons, primarily through power-law and logarithmic corrections. Logarithmic entropy, such as Shannon entropy, is crucial in quantum mechanics. However, in non-asymptotic regimes where the law of large numbers is less applicable, alternative models like collision entropy gain prominence.
To address these issues, Rényi introduced a generalized entropy framework that unifies various entropy measures and allows modifications to the black hole area law. Rényi entropy generalizes Boltzmann-Gibbs entropy by incorporating a tunable parameter that adjusts the sensitivity to different probabilities in the system. This flexibility is particularly useful in scenarios where rare or highly probable events require distinct emphasis. Moreover, Rényi entropy is closely related to quantum entanglement, describing situations where the quantum states of multiple particles become interlinked, thereby providing a valuable tool for studying complex quantum systems.\\
The Rényi entropy is expressed as \cite{Reny} :
\begin{equation}
    S_{R} = \frac{1}{1-q} \left(\ln \sum_{i=1}^{n} P^{q}(i) \right),
\end{equation}
where \( P(i) \) denotes the probability distribution and \( q \) is the non-extensive Tsallis parameter. In this work, we assume that the black hole entropy follows the Tsallis entropy framework \cite{Tsallis}, which generalizes the Boltzmann-Gibbs entropy through a single parameter, thereby encompassing both extensive and non-extensive statistical systems. The Tsallis entropy is given by\cite{Tsallis}:
\begin{equation}
    S_T = S_{BH} = \frac{1}{1-q} \left( \sum_{i=1}^{n} P^{q}(i) - 1 \right).
\end{equation}
In the limit \( q \to 1 \), Tsallis entropy reduces to the standard Boltzmann-Gibbs entropy. The parameter \( q \) governs the degree of non-extensivity, with distinct behaviors depending on its value:
\begin{itemize}
    \item \textbf{For \( q = 1 \)}: Tsallis entropy coincides with the extensive Boltzmann-Gibbs entropy, suitable for systems with weak correlations and short-range interactions.
    \item \textbf{For \( q < 1 \)}: The entropy exhibits sub-extensive behavior, applicable to systems where rare states occur with a higher probability than frequent states.
    \item \textbf{For \( q > 1 \)}: The entropy is super-extensive, where frequent states contribute more significantly to the total entropy.
\end{itemize}
Substituting into the general form, the Rényi entropy can be re-expressed as:
ole area law entropy.The expression for Renyi entropy is given by  \cite{Reny} 
	\begin{equation}
	S=\frac{1}{\alpha} ~ ln[1+\alpha S_0]
	\label{renyi}
	\end{equation}
	Where $S_0$ is the Bekenstein entropy for black holes. $\alpha=1-q$ is the  Renyi
parameter. When  $\alpha$ goes to $0$,  we obtain the Bekenstein entropy again.\\

Since its formulation in 1915, General Relativity (GR) has stood as the foundation of modern theoretical physics. Recognized as a highly successful theory of gravity, GR has been validated through various experimental observations, including the perihelion precession of Mercury, the bending of light during the 1919 solar eclipse, the groundbreaking detection of gravitational waves by the Laser Interferometer Gravitational-Wave Observatory (LIGO) in 2015 \cite{ligo}, and the remarkable 2019 imaging of the supermassive black hole at the center of galaxy M87 by the Event Horizon Telescope (EHT) \cite{m87a, m87b, m87c, m87d, m87e, m87f}. Despite these triumphs, GR faces considerable challenges, particularly with the discovery of the universe's accelerating expansion \cite{reiss, perlmutter, spergel, astier} and the anomalies in galaxy rotation curves, suggesting the presence of an unseen form of matter \cite{naselskii}, commonly known as dark matter.\\

The pursuit of quantum gravity effects has garnered significant attention in recent decades. On the experimental front, substantial progress has been made. Theoretically, the investigation of Lorentz symmetry breaking (LSB) plays a crucial role in advancing our understanding of quantum gravity in fundamental physics\cite{Altschul2010,Kalb1974,Kao1996,Kar2003,Chakraborty2017}. By examining how Lorentz symmetry breaking contributes at low energy scales, particularly its effects on spacetime, and by analyzing high-precision experimental data, we can investigate its compatibility with General Relativity (GR).As a result, Lorentz symmetry breaking (LSB) has emerged as a prominent focus in black hole physics research.\\

An important model that incorporates Lorentz symmetry violation is the Kalb-Ramond (KR) gravity theory, which introduces a tensor field $B_{ab}$ through non-minimal coupling. This KR field originates from the bosonic sector of string theory \cite{kr_bosonic_1,kr_bosonic_2}. The field $B_{ab}$ is an antisymmetric second-rank tensor that can be formulated as:  
\begin{equation}
B_{ab} = \tilde{E}_{[a}v_{b]} + \epsilon_{abcd}v^c \tilde{B}^d
\end{equation}
Here, $v^a$ is a timelike four-vector, while $\tilde{E}^a$ and $\tilde{B}^a$ are spacelike pseudoelectric and pseudomagnetic fields, respectively. The constraints $\tilde{E}^a v_a = 0$ and $\tilde{B}^a v_a = 0$ ensure that both fields are orthogonal to $v^a$. Although analogous to the electric and magnetic fields in Maxwell's theory, these fields arise specifically within string theory. \\

Research in this field has primarily aimed at obtaining exact solutions to the Einstein-Kalb-Ramond field equations \cite{kr_exact_1, kr_exact_2, main, kr_exact_4}. A significant advancement was achieved by Yang et al. \cite{kr_exact_5}, who formulated a Schwarzschild-like solution under this framework. By relaxing the vacuum constraints, this approach was later extended to yield a Schwarzschild-(A)dS-like solution. Recently, interest has surged in exploring the spacetime characteristics of these black hole solutions.  
Despite these developments, the existing solutions have been confined to a specific spherically symmetric configuration with $-g_{tt} = g^{-1}_{rr}$. Broader cases where $-g_{tt} \neq g^{-1}_{rr}$ have yet to be thoroughly investigated. Recent studies have explored various aspects of Kalb-Ramond (K-R) gravity\cite{kr1,kr2,kr3,kr4}.In ref. \cite{main}, authors presented an exact solutions for
static and spherically symmetric black holes in the framework of this Lorentz-violating gravity theory.  In our work, we have consider that particular solution.\\

A recent advancement in this field is the topological perspective on black hole phase transitions \cite{Wei:2022dzw,Wei:2021vdx,Yerra:2022alz,Yerra:2022eov,Gogoi:2023xzy,Gogoi:2023qku,Gogoi:2023wih,Yerra:2022coh,Yerra:2023ocu,Barzi:2023msl,yerrabm,Ahmed:2022kyv,Wei:2022mzv,Fan:2022bsq,Wu:2022whe,Fang:2022rsb,Wu:2023xpq,Wu:2023sue,Li:2023ppc,Wei:2023bgp,Alipour:2023uzo,Zhang:2023uay,Sadeghi:2023aii,Wang:2024zbp,Shahzad:2024ojx,Malik:2024kau,Zhao:2024tlu,Wu:2024rmv,Hazarika:2024cpg,Hazarika:2023iwp,Sadeghi:2024krq,Zhang:2023svu,Bhattacharya:2024bjp,J,nwu1,nwu2,nwu3,nwu5,nwu4}, where critical points are associated with specific topological charges. Several studies have employed thermodynamic topology to explore van der Waals-type transitions \cite{Wei:2021vdx,Yerra:2022alz,Yerra:2022eov,Gogoi:2023xzy,Gogoi:2023qku}, Hawking-Page transitions \cite{Yerra:2022coh,Yerra:2023ocu,Barzi:2023msl,yerrabm}, and Davies-type transitions \cite{Bhattacharya:2024bjp}. The literature indicates that studying Hawking-Page and Davies-type phase transitions from a topological viewpoint often involves defining distinct potentials or vector fields \cite{Zhang:2023uay,Yerra:2022coh,Bhattacharya:2024bjp,Fan:2022bsq}.
The foundation of black hole thermodynamic topology is inspired by Duan’s $\phi$-mapping current theory \cite{Duan,Duan:2018rbd}. The most reliable method to analyze thermodynamic topology of a black hole system was provided in the ref.\cite{Wei:2022dzw}.In this work we have adopted that particular method.\\

In this paragraph we have presented a brief overview of the Duan's $\phi$-mapping  method to calculate winding number and topological charge for any two dimensional vector field.
In Duan’s formalism, a vector field $\phi = \{\phi^a\}$ (where $a = 1,2$) in the coordinate space $x^\nu = \{t, r_+, \theta\}$ can be used to define a topological current \cite{Duan}:
\begin{equation}
    j^\mu = \frac{1}{2\pi} \epsilon^{\mu \nu \rho} \epsilon_{ab} \partial_\nu n^a \partial_\rho n^b,
\end{equation}
where $\partial_\nu = \frac{\partial}{\partial x^\nu}$, and $\mu, \nu, \rho = 0, 1, 2$. The normalized vector $n^a$ is defined as:
\begin{equation}
    n^a = \frac{\phi^a}{||\phi||}, \quad a = 1, 2, \quad \text{with} \quad \phi^1 = \phi^{r_+}, \quad \phi^2 = \phi^{\theta}.
\end{equation}
The normalized vector $n^a$ satisfies the following conditions \cite{Duan}:
\begin{equation}
    n^a n^a = 1, \quad n^a \partial_\nu n^a = 0.
\end{equation}
It is straightforward to verify the conservation of the topological current:
\begin{equation}
    \partial_\mu j^\mu = 0.
\end{equation}
Using the Jacobi tensor $\epsilon^{ab} J^\mu \left(\frac{\phi}{x} \right) = \epsilon^{\mu \nu \rho} \partial_\nu \phi^a \partial_\rho \phi^b$ and the two-dimensional Laplacian Green function $\Delta_{\phi^a} \ln ||\phi|| = 2\pi \delta^2(\phi)$, the topological current can be expressed as \cite{Duan}:
\begin{equation}
    j^\mu = \delta^2(\phi) J^\mu \left( \frac{\phi}{x} \right).
\end{equation}
This implies that $j^\mu$ is non-zero only at points where $\phi^a(x^i) = 0$, denoted as $\vec{x} = \vec{z}_i$. By applying the $\delta$-function theory \cite{schouten}, the topological current density can be written as:
\begin{equation}
    j^0 = \sum_{i=1}^N \beta_i \eta_i \delta^2(\vec{x} - \vec{z}_i).
\end{equation}
Here, $\beta_i$ is the Hopf index, representing the number of loops formed by $\phi^a$ as $x^\mu$ encircles the zero point $z_i$, and $\eta_i = \text{sign}\left( J^0(\phi/x)_{z_i} \right) = \pm 1$ denotes the Brouwer degree. The topological charge within a region $\Sigma$ is then calculated as:
\begin{equation}
    Q = \int_\Sigma j^0 d^2x = \sum_{i=1}^N \beta_i \eta_i = \sum_{i=1}^N w_i,
\end{equation}
where $w_i$ denotes the winding number of the $i$-th zero of $\phi$.\\

Recent research highlights that flat black holes under the Rényi entropy framework display thermodynamic behavior closely resembling that of AdS black holes in the Bekenstein entropy paradigm. This resemblance was initially investigated in \cite{Barzi1} and later expanded upon in various studies \cite{Barzi2,proof3,proof4,proof5,proof6,proof7,proof8,proof9}. Notably, \cite{expanding} demonstrated that the Friedmann equations can be derived using Rényi entropy, allowing the cosmological constant \(\Lambda\) to be expressed in terms of the Rényi parameter \(\alpha\). This approach removes the need for \(\Lambda\) to be manually introduced into the Einstein–Hilbert action.\\

These similarities have been observed within the framework of Einstein's general theory of relativity. This naturally raises the question: do they extend to other theories of gravity? Motivated by this inquiry, we propose exploring a potential correspondence from a topological perspective between the thermodynamics of black holes in asymptotically flat spacetimes governed by Rényi statistics and those in asymptotically Anti-de-Sitter (AdS) spacetimes described by Gibbs-Boltzmann statistics, specifically within the context of Kalb-Ramond (KR) gravity theory.\\

\section{Static Flat Black Hole in K-R Gravity}
\begin{figure}[h]
    \centering
    \begin{subfigure}[t]{0.32\textwidth}
        \centering
        \includegraphics[width=\linewidth]{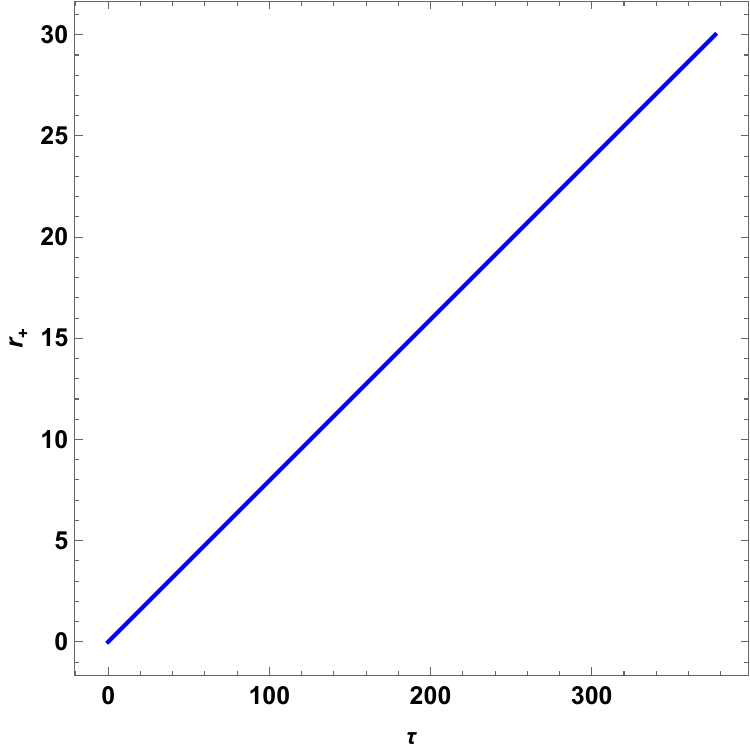}
        \caption{a}
        \label{rf1a}
    \end{subfigure}
    \hfill
    \begin{subfigure}[t]{0.32\textwidth}
        \centering
        \includegraphics[width=\linewidth]{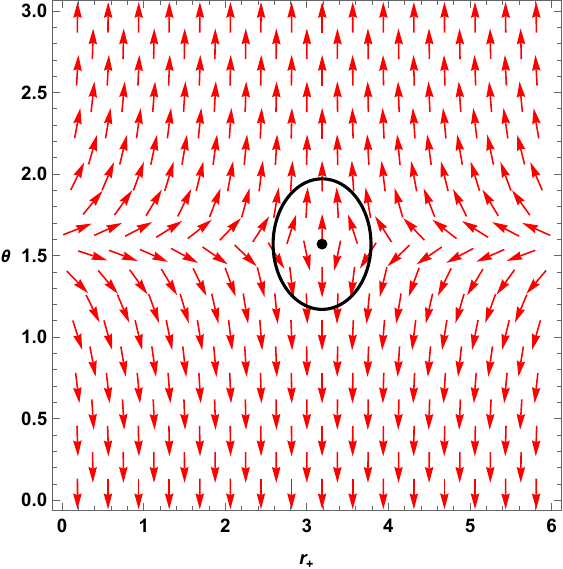}
        \caption{b}
        \label{rf1b}
    \end{subfigure}
    \hfill
    \begin{subfigure}[t]{0.32\textwidth}
        \centering
        \includegraphics[width=\linewidth]{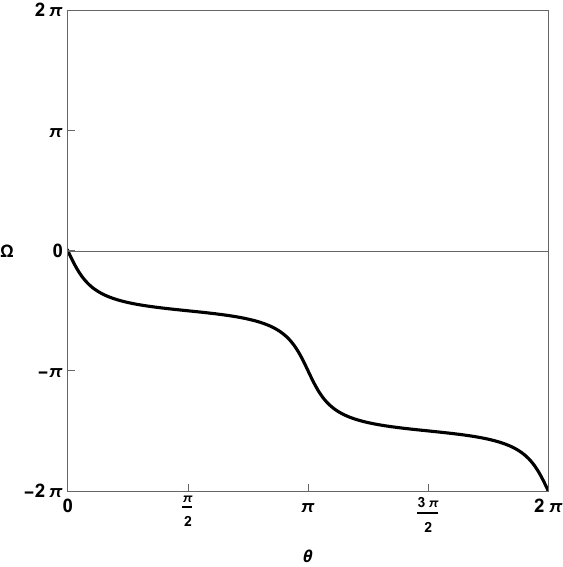}
        \caption{c}
        \label{rf1c}
    \end{subfigure}
    \caption{Winding Number. $\ell$=0.001, $\tau$=40}
    \label{rf1}
\end{figure}

In this section, we study the thermodynamic topology of static flat black holes in Kalb-Ramond Gravity. 
The general static neutral spherically symmetric black hole solution in K-R gravity is obtained as \cite{main} :
\begin{equation}\label{dscaseA}
\begin{aligned}
ds^2=&-\left(\frac{1}{1-\ell}-\frac{2 M}{r}\right)dt^2+\frac{1}{\left(\frac{1}{1-\ell}-\frac{2 M}{r}\right)}dr^2
&+r^2d\theta^2+r^2\sin^2\theta d\varphi^2.
\end{aligned}
\end{equation}
Where $\ell$ is a  dimensionless constant known as the Lorentz-violating parameter. It can have both positive and negative value and the value is usually kept very small.In ref.\cite{main}, the constraints on $\ell$ is obtained as $-3.7 \times 10^{-12} \leq \ell \leq 1.9 \times 10^{-11}$.
The mass of the black hole is calculated by setting the metric function equal to zero at event horizon radius $r_+$ as 
\begin{equation}
M=\frac{r_+}{2 (1-\ell)}
\end{equation}
Using the metric function, the temperature of the black hole at the event horizon is calculated to be
\begin{equation}
T=\frac{1}{4 \pi  (1-\ell) r_+}
\end{equation}
From the expression for mass and temperature, the entropy of the black hole is calculated using the following formula :
\begin{equation}
S_0=\int \frac{d M}{T}=\pi r_{+}^2
\label{entropy}
\end{equation}
First,  we shall perform our topological analysis in the usual Gibbs-Boltzmann statistics. We start off by writing down the expression for the off-shell free energy\cite{Wei:2022dzw}, 
\begin{equation}\label{r1}
\mathcal{F}=M-\frac{S}{\tau}=r_+ \left(\frac{1}{2-2 \ell}-\frac{\pi  r_+}{\tau }\right)
\end{equation}
Using the off-shell free energy, a two dimensional vector field in $r-\theta$ plan can be constructed as \cite{Wei:2022dzw} :
\begin{equation}
(\phi^r,\phi^\theta )=\biggl(\frac{\partial \mathcal{F}}{\partial r}, -cot \theta \csc \theta \biggr)
\end{equation}
The $\phi^r$ component can be calculated by taking the first-order derivative of the free energy,
\begin{equation}\label{r2}
\phi^r=\frac{1}{2-2 \ell}-\frac{2 \pi  r_+}{\tau }
\end{equation}
 We solve $\phi^r=0$ to get obtain the value for $\tau$ which comes out to be
\begin{equation}\label{r3}
\tau=-4 \pi  (\ell-1) r_+
\end{equation}
The $\tau$ vs $r_+$ plot is shown in Figure \ref{rf1a}. It is evidently just a straight line showing a single branch of the black hole. Figure \ref{rf1b} shows the vector plot of $\left(\phi^r , \phi^\Theta\right)$ in the $r_+ - \theta$ plane. In this case, the zero point for $\tau=40$ is observed at $r_+=3.18629$. The winding number can be calculated by using the Duan's $\phi$- mapping technique as we have discussed earlier. The winding number corresponding to $r_+=3.18629$ is $-1$ as can be inferred from Figure \ref{rf1c} and accordingly, the topological charge will be $-1$.

\section{static flat black hole in K-r gravity using Renyi statistics}
In this section, we study the thermodynamic topology of the static flat black holes in Kalb-Ramond gravity while incorporating Renyi statistics. We first write the horizon radius $r_+$ in terms of the Renyi entropy $S$ using eq.(\ref{renyi}) and eq.(\ref{entropy}) as
\begin{equation}\label{r4}
r_+=\frac{\sqrt{e^{\alpha  S}-1}}{\sqrt{\pi } \sqrt{\alpha }}
\end{equation}
Using this equation the expression of mass can be rewritten in terms of the Renyi entropy as
\begin{equation}\label{r5}
M_R=\frac{\sqrt{e^{\alpha  S}-1}}{\sqrt{\pi } \sqrt{\alpha } (2-2 \ell)}
\end{equation}
Temperature in the Renyi entropy framework is calculated as
\begin{equation}
T_R=\frac{d M_R}{d S}=\frac{\sqrt{\alpha } e^{\alpha  S}}{\sqrt{\pi } (4-4 \ell) \sqrt{e^{\alpha  S}-1}}
\end{equation}
Therefore the expression for free energy will be modified to
 \begin{equation}\label{r6}
\mathcal{F}=M-\frac{S}{\tau}=\frac{\sqrt{e^{\alpha  S}-1}}{\sqrt{\pi } \sqrt{\alpha } (2-2 \ell)}-\frac{S}{\tau }
\end{equation}
The $\phi^r$ component can be calculated from the free energy as before
\begin{equation}\label{r7}
\phi^r=\frac{\sqrt{\alpha } e^{\alpha  S}}{\sqrt{\pi } (4-4 \ell) \sqrt{e^{\alpha  S}-1}}-\frac{1}{\tau }
\end{equation}
Solving $\phi^r=0$ we find out the expression of $\tau$
\begin{equation}\label{r8}
\tau=-\frac{4 \sqrt{\pi } (\ell-1) e^{\alpha  (-S)} \sqrt{e^{\alpha  S}-1}}{\sqrt{\alpha }}
\end{equation}

\begin{figure}[h]
    \centering
    \begin{subfigure}[t]{0.32\textwidth}
        \centering
        \includegraphics[width=\linewidth]{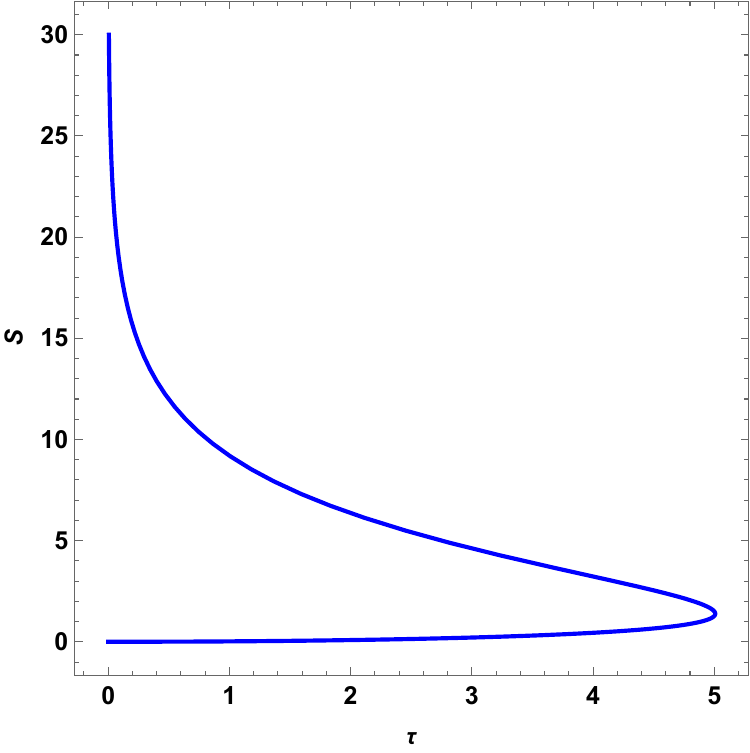}
        \caption{a}
        \label{rf2a}
    \end{subfigure}
    \hfill
    \begin{subfigure}[t]{0.32\textwidth}
        \centering
        \includegraphics[width=\linewidth]{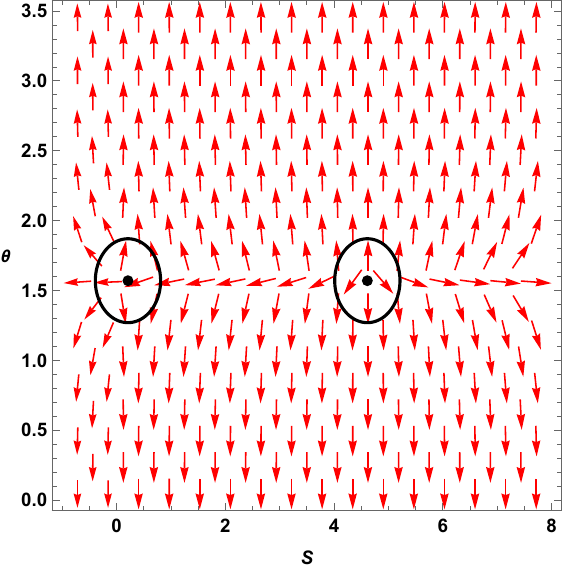}
        \caption{b}
        \label{rf2b}
    \end{subfigure}
    \hfill
    \begin{subfigure}[t]{0.32\textwidth}
        \centering
        \includegraphics[width=\linewidth]{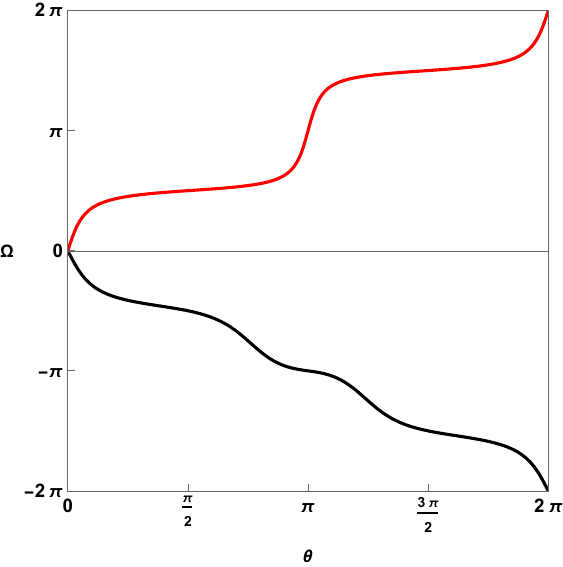}
        \caption{c}
        \label{rf2c}
    \end{subfigure}
    \caption{Winding Number. $\ell=0.001$, $\alpha=0.5$, $\tau=3$}
    \label{rf2}
\end{figure}

 We show the $\tau$ vs $S$ plot in Figure \ref{rf2a}. It can be observed that there are two black hole branches, one is the small black hole branch (with lower entropy) and the other is the large black hole branch (with higher entropy). The vector plot in the $S-\theta$ plane in Figure \ref{rf2b} has its vanishing points at $S=0.209899$ and $S=4.61258$ for an arbitrary value of $\tau=3$. From Figure \ref{rf2c}, we can conclude that the winding number corresponding to $S=0.209899$ and $S=4.61258$ are $-1$ and $+1$ respectively. Hence the topological charge, $W$ will be zero. The winding number reveals the stability of the black hole branches. For the large black hole branch, the winding number is positive, indicating local stability, while for the small black hole branch, it takes a negative value, suggesting local instability. The precise transition point between the small and large black hole branches is found to be at $(\tau, S) = (5.00824, 1.38629)$. This critical point is identified as a generation point, where an unstable black hole branch ends and a stable branch begins to emerge. From this, we can draw a significant conclusion: the static black holes in GB statistics and Renyi statistics are topologically distinct. Another fascinating observation arises from the differing behaviors between the two frameworks: while no phase transition is detected in the GB statistics, both Hawking-Page and Davies-type phase transitions are observed within the Renyi statistics.\\

 The topology of the phase transition point can also be calculated by employing a method shown in ref \cite{J}. For that, first, we write down the expression for  free energy(not off-shell) of the black hole as
\begin{equation}
F=M_R-T_R S=\frac{e^{\alpha  S} (\alpha  S-2)+2}{4 \sqrt{\pi } \sqrt{\alpha } (l-1) \sqrt{e^{\alpha  S}-1}}
\end{equation}
Utilizing this free energy expression, a new vector field $\Phi$ in $S-\theta$ plane is constructed as \cite{J} :
\begin{equation}
\Phi=(\phi^S, \phi^\theta)=\biggl( \frac{d F^2}{d S}, -\cot \theta \csc \theta    \biggr)
\end{equation}
By performing some simple mathematical steps, it becomes evident that there are two zero points of the vector component $\phi^S$. One zero point occurs when $F = 0$, and the other occurs when $\frac{1}{C}$ tends to zero, where $C = \frac{dM}{dT}$ represents the heat capacity of the black hole. The condition $F = 0$ corresponds to the Hawking-Page phase transition point, while $\frac{1}{C}$ corresponds to the Davies point.\\

\begin{figure}
\includegraphics[scale=0.5]{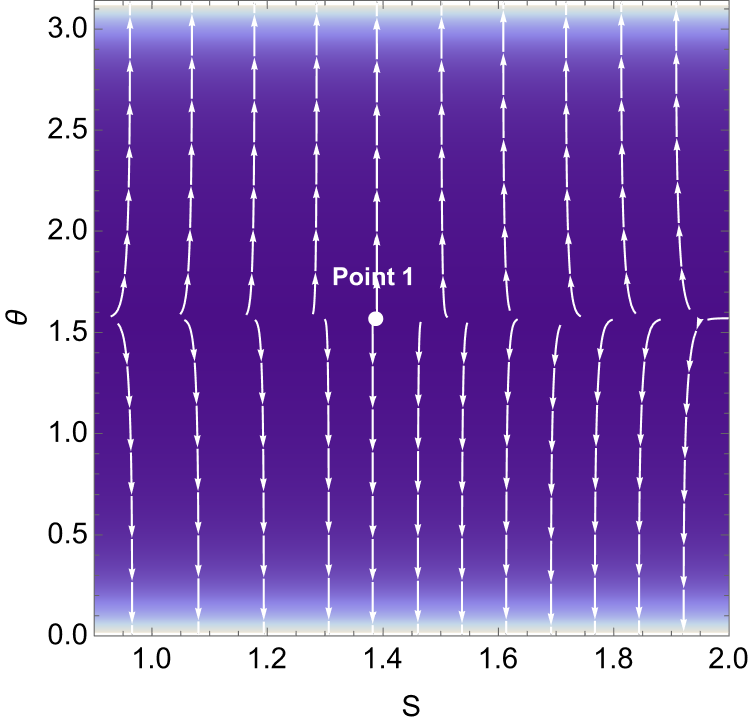}
\hspace{0.5cm}
\includegraphics[scale=0.5]{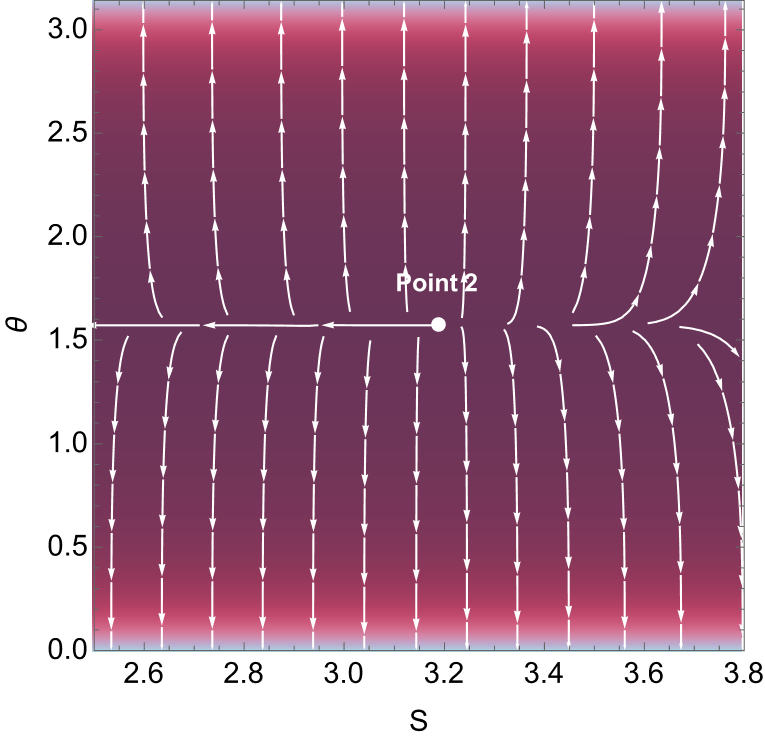}
\caption{Zero points of the vector field eq.(\ref{comp}). "Point 1" is the Davies point and  "Point 2" is the Hawking-Page phase transition point. }
\label{vec}
\end{figure}

The $ \phi^S$ component in this particular scenario is calculated to be
\begin{equation}
\begin{aligned}
\phi^S &= \frac{\alpha  S e^{\alpha  S} \left(e^{\alpha  S}-2\right) \left(e^{\alpha  S} (\alpha  S-2)+2\right)}{16 \pi  (\ell-1)^2 \left(e^{\alpha  S}-1\right)^2} \\
\phi^\theta &= -\cot\theta\csc\theta
\end{aligned}
\label{comp}
\end{equation}
We plot the vector components in the $S-\theta$ plane with $\ell = 0.01$ and $\alpha = 0.5$, as shown in Figure \ref{vec}. Two zero points are observed in this plot. "Point 1" corresponds to the zero point where $F = 0$, and is located at $S = 1.38629$. The second zero point, "Point 2", corresponds to $1/C = 0$, and is identified at $S = 3.18724$.

\begin{figure}[h!]
\includegraphics[scale=0.5]{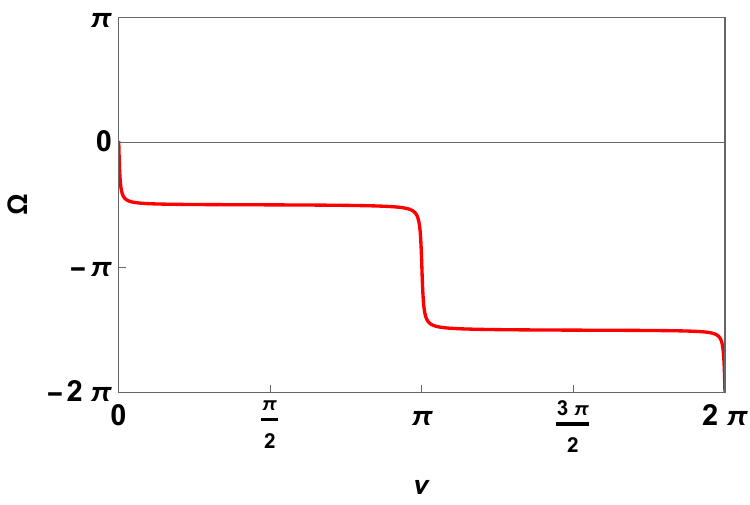}
\hspace{0.5cm}
\includegraphics[scale=0.5]{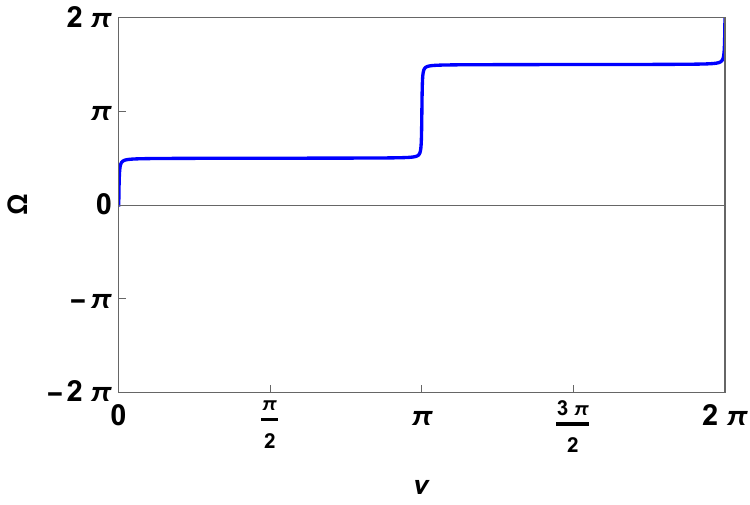}
\caption{Calculation of topological charge by plotting deflection around "Point 1" and "Point 2." The red contour shows the deflection for Davies point and the blue contour represents the same for Hawking-Page phase transition point. }
\label{contour}
\end{figure}

The topological charge of the zero points is determined using Duan's $\phi$-mapping technique, as illustrated in Figure. \ref{contour}. The analysis reveals that the deflection around the Davies point is negative, resulting in a topological charge of $-1$, which is indicated by the red-colored line. In contrast, the deflection at the Hawking-Page phase transition point is positive, leading to a topological charge of $+1$, as represented by the blue-colored line.\\

Hence, we conclude that in the framework of GB statistics, the static flat black hole in K-R gravity possesses a topological charge of $-1$ with no phase transition observed. In contrast, within the framework of Renyi entropy, the topological charge is changed to $0$, and the black hole exhibits both Hawking-Page and Davies-type phase transitions. Additionally, we identify a generation point in this case. The topological charge at the critical point is found to be $+1$ for the Hawking-Page phase transition point and $-1$ for the Davies-type phase transition point.

\section{AdS Black Hole in K-R Gravity}
Next, we study the thermodynamic topology of AdS black holes in K-R gravity in GB statistics. We take exactly the same approaches adopted in the previous sections. 
The corresponding (A)dS extension  of the static K-R black hole is \cite{main}
\begin{equation}
\begin{aligned}
\label{KRAdS}
ds^2 = & -\left[\frac{1}{1-\ell} - \frac{2M}{r} - \frac{\Lambda r^2}{3(1-\ell)}\right] dt^2 
& + \frac{dr^2}{\frac{1}{1-\ell} - \frac{2M}{r} - \frac{\Lambda r^2}{3(1-\ell)}} + r^2 d\theta^2 + r^2 \sin^2\theta \, d\varphi^2.
\end{aligned}
\end{equation}

In AdS space, the cosmological constant is written in terms of the AdS boundary $\kappa$ as $\Lambda=-\frac{3}{\kappa^2}$. The mass of the AdS black hole is evaluated to be
\begin{equation}
M=\frac{r_+ \left(\kappa ^2+r_+^2\right)}{2 \kappa ^2 (1-\ell)}
\end{equation}

The entropy of the AdS black hole is 
\begin{equation}
S_0=\pi r_{+}^2
\end{equation}

The off-shell free energy comes out to be \begin{equation}\label{r9}
\mathcal{F}=M-\frac{S}{\tau}=-\frac{r_+ \left(\ell^2+r_+^2\right)}{2 (\kappa -1) l^2}-\frac{\pi  r_+^2}{\tau }
\end{equation}
Taking the first derivative, we compute the $\phi^r$ component as 
\begin{equation}\label{r10}
\phi^r=-\frac{l^2+3 r_+^2}{2 (\kappa -1) \ell^2}-\frac{2 \pi  r_+}{\tau }
\end{equation}
By equating $\phi^r$ to $0$ we get the expression for $\tau$ as
\begin{equation}\label{r11}
\tau=-\frac{4 \pi  \left(\kappa \ell^2 r_+-\ell^2 r_+\right)}{\ell^2+3 r_+^2}
\end{equation}

\begin{figure}[h!]
    \centering
    \begin{subfigure}[t]{0.32\textwidth}
        \centering
        \includegraphics[width=\linewidth]{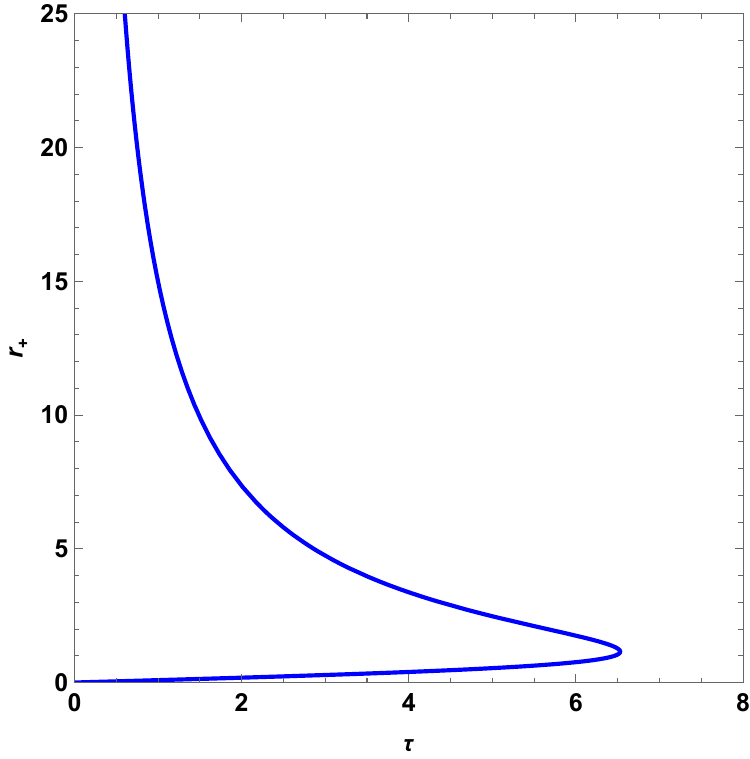}
        \caption{a}
        \label{rf3a}
    \end{subfigure}
    \hfill
    \begin{subfigure}[t]{0.32\textwidth}
        \centering
        \includegraphics[width=\linewidth]{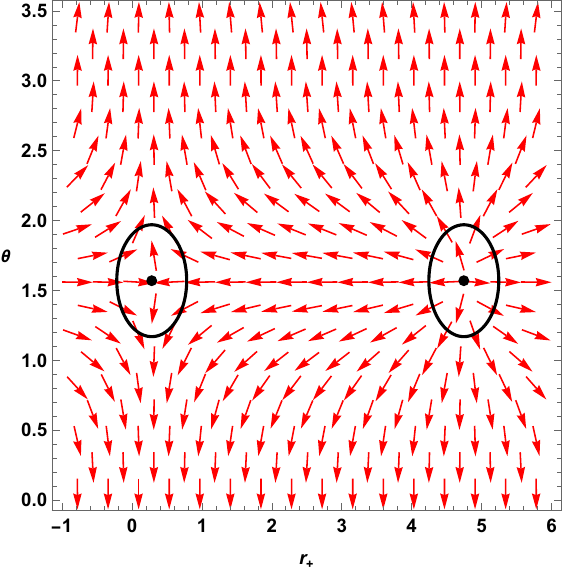}
        \caption{b}
        \label{rf3b}
    \end{subfigure}
    \hfill
    \begin{subfigure}[t]{0.32\textwidth}
        \centering
        \includegraphics[width=\linewidth]{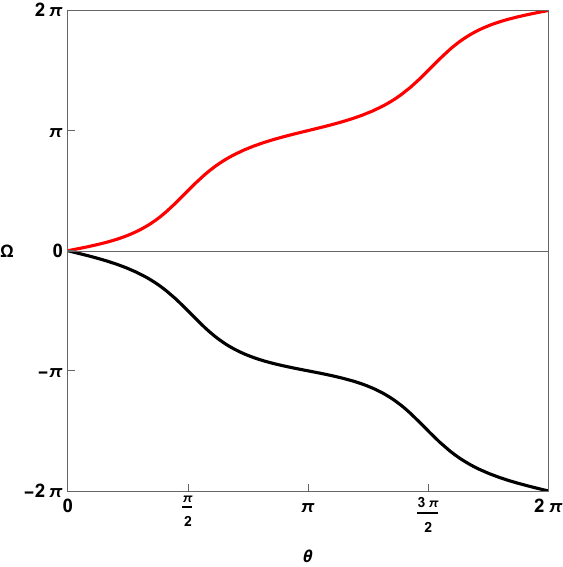}
        \caption{c}
        \label{rf3c}
    \end{subfigure}
    \caption{Winding Number. $\kappa=0.1$, $\ell=2$, $\tau=3$}
    \label{rf3}
\end{figure}

The plot of $\tau$ vs $r_+$ in Figure \ref{rf3a}. We observe that in this case also there are two black hole branches similar to the previous case. The vector plot in the $r_+-\theta$ plane is shown in Figure \ref{rf3b}. It has two vanishing points at $r_+=0.280963$ and $r_+=4.74559$ for any arbitrary value of $\tau=3$. The winding numbers corresponding to $r_+=0.280963$ and $r_+=4.74559$ are respectively $-1$ and $+1$ as can be seen from Figure \ref{rf3c}. Therefore the topological charge for AdS black holes in K-R gravity is $W=0$.In this scenario, we also identified a generation point, marking the emergence of a new stable black hole branch. Furthermore, the presence of both Hawking-Page and Davies-type phase transitions was observed. By employing the method discussed earlier, we determined the topological charge associated with the Hawking-Page phase transition to be $+1$, while the topological charge corresponding to the Davies-type phase transition was found to be $-1$.  Thus, we conclude that the global and local topological properties of the K-R AdS black hole in the Gibbs-Boltzmann statistical framework are equivalent to those of the K-R flat black hole within the Renyi statistical framework.

\section{Tentative connection Between Renyi Parameter and Cosmological Constant}
From the results of the previous section, it is evident that the K-R flat black hole in Renyi statistics is topologically equivalent to the K-R AdS black hole in GB statistics. This observation raises an obvious question: Is there a connection between the Renyi entropy parameter and the cosmological constant? Specifically, in the case of the flat black hole, could the absence of the cosmological constant be compensated by the inclusion of the Renyi parameter? In this section, we aim to explore this possibility and establish a mathematical relationship between the Renyi entropy parameter and the cosmological constant.
To begin, let us consider the mass of a K-R flat black hole in Renyi statistics frame work,
\begin{equation}
M_R=\frac{\sqrt{e^{\alpha  S}-1}}{\sqrt{\pi } \sqrt{\alpha } (2-2 \ell)}
\label{masssc1}
\end{equation}
	Assuming $\alpha$ to be very small, we can expand Eq.~(\ref{masssc1}) up to the first order while neglecting higher-order terms. This expansion yields:  
\begin{equation}
M_R =  - \frac{\sqrt{S}}{2 \sqrt{\pi} (\ell-1)}-\frac{\alpha S^{3/2}}{8 \sqrt{\pi} (\ell-1)} + O(\alpha^{3/2}).  
\label{series}
\end{equation}  

Next, we examine the mass of the K-R AdS black hole within the GB statistics framework, which is expressed as:  
\begin{equation}
M =  - \frac{\sqrt{S_0}}{2 \sqrt{\pi} (\ell-1)}-\frac{S_0^{3/2}}{2 \pi^{3/2} \kappa^2 (\ell-1)},  
\label{massads}
\end{equation}  
where $S_0 = \pi r_+^2$ represents the black hole entropy.  

By comparing Eq.~(\ref{massads}) with the first two terms of the series expansion in Eq.~(\ref{series}), we deduce:  
\begin{equation}
\alpha \approx \frac{4}{\pi \kappa^2}.  
\label{alpha}
\end{equation}  

Substituting $\kappa^2 = -\frac{3}{\Lambda}$ into Eq.~(\ref{alpha}), we find:  
\begin{equation}
\alpha \approx -\frac{4 \Lambda}{3 \pi},  
\end{equation}  
or equivalently,  
\begin{equation}
\Lambda \approx -\frac{3 \pi \alpha}{4}.  
\end{equation}  

A comparable relationship was derived in Ref.~\cite{expanding}, where the author investigated the corrections to the Friedmann equations within the framework of Rényi entropy. By analyzing the late-time evolution using a first-order perturbative approach, the study presented a cosmological model that establishes a connection between the Rényi parameter $\alpha$ and the cosmological constant $\Lambda$ as :
\begin{equation}
\Lambda \approx \pm 3 \pi \alpha.  
\label{ref1}
\end{equation}  

From Eqs.~(\ref{alpha}) and (\ref{ref1}), it becomes evident that a link exists between the AdS length scale $\kappa$ and the entropy parameter $\alpha$. Specifically, $\alpha$ is found to scale as $\alpha \propto \frac{1}{\kappa^2}$.

	\section{Conclusion}

In this work, we have explored the topological and thermodynamic properties of K-R black holes in both flat and AdS spacetimes, within the frameworks of GB and Renyi statistics. Our analysis reveals significant differences in the topological characteristics and phase transition behavior between these scenarios. First, we compared the topology of K-R flat black holes in GB and Renyi statistical frameworks. The transition from GB to Renyi statistics introduced notable changes: the total topological charge shifted from $W = -1 $ to $ W = 0$, indicating a topological transformation. Additionally, we identified a generation point within the Renyi framework, a feature absent in the GB statistical framework. The Renyi framework also allowed for the observation of both Hawking-Page and Davies-type phase transitions, which were not present in the GB framework. The topological charge associated with the Hawking-Page phase transition point was calculated to be $ +1 $, while that of the Davies-type phase transition point was found to be $ -1$. Next, we examined the topology of K-R AdS black holes within the GB statistical framework. Here, we observed a total topological charge of $ W = 0 $, along with the presence of a generation point marking the emergence of a new stable black hole branch. As in the Renyi framework for flat black holes, we identified both Hawking-Page and Davies-type phase transitions. The corresponding topological charges were found to be $+1 $ for the Hawking-Page phase transition point and $ -1 $ for the Davies-type phase transition point. From these results, we conclude that the global and local topological properties of the K-R AdS black hole in GB statistics are equivalent to those of the K-R flat black hole in Renyi statistics. This equivalence suggests a deeper connection between the cosmological constant and the Renyi parameter. Through our analysis, we derived an approximate relationship between the Renyi parameter and the cosmological constant, which aligns with similar findings in the literature derived from cosmological considerations. These similarities could either stem from purely mathematical analogies or hint at a deeper connection between the Rényi entropy and the cosmological constant. The underlying reason for this resemblance remains an open question, warranting further investigation. Should a direct link between the cosmological constant and Rényi entropy exist, as indicated by our findings and supported by recent studies, it may provide valuable insights into several unresolved issues in black hole physics and cosmology.

\section{Acknowledgments}
BH would like to thank DST-INSPIRE, Ministry of Science and Technology fellowship program, Govt. of India for awarding the DST/INSPIRE Fellowship[IF220255] for financial support. 	

	\bibliographystyle{apsrev}

\begin{thebibliography}{99}
		
		\bibitem{Bekenstein:1973ur}
		J.~D.~Bekenstein,
		Black holes and entropy,
		Phys. Rev. D \textbf{7}, 2333-2346 (1973)
		doi:10.1103/PhysRevD.7.2333

		\bibitem{Hawking:1974rv}
		S.~W.~Hawking,
		Black hole explosions,
		Nature \textbf{248}, 30-31 (1974)
		doi:10.1038/248030a0
		\bibitem{Hawking:1975vcx}
		S.~W.~Hawking,
		Particle Creation by Black Holes,
		Commun. Math. Phys. \textbf{43}, 199-220 (1975)
		[erratum: Commun. Math. Phys. \textbf{46}, 206 (1976)]
		doi:10.1007/BF02345020
		
		
		
		
		
		
		
		
		\bibitem{Bardeen:1973gs}
		J.~M.~Bardeen, B.~Carter and S.~W.~Hawking,
		The Four laws of black hole mechanics,
		Commun. Math. Phys. \textbf{31}, 161-170 (1973)
		doi:10.1007/BF01645742
		
		
		
		
		
		
		\bibitem{Wald:1979zz}
		R.~M.~Wald,
		Entropy and black-hole thermodynamics,
		Phys. Rev. D \textbf{20}, 1271-1282 (1979)
		doi:10.1103/PhysRevD.20.1271
		\bibitem{bekenstein1980black}
		Jacob~D Bekenstein.
		Black-hole thermodynamics,
		Physics Today, 33(1):24--31, 1980.
		\bibitem{Wald:1999vt}
		R.~M.~Wald,
		The thermodynamics of black holes,
		Living Rev. Rel. \textbf{4}, 6 (2001)
		doi:10.12942/lrr-2001-6
		[arXiv:gr-qc/9912119 [gr-qc]].
		
		\bibitem{Carlip:2014pma}
		S.Carlip,
		Black Hole Thermodynamics,
		Int. J. Mod. Phys. D \textbf{23}, 1430023 (2014)
		doi:10.1142/S0218271814300237
		[arXiv:1410.1486 [gr-qc]].
		\bibitem{Wall:2018ydq}
		A.C.Wall,
		A Survey of Black Hole Thermodynamics,
		[arXiv:1804.10610 [gr-qc]].
		\bibitem{Candelas:1977zz}
		P.Candelas and D.W.Sciama,
		Irreversible Thermodynamics of Black Holes,
		Phys. Rev. Lett. \textbf{38}, 1372-1375 (1977)
		doi:10.1103/PhysRevLett.38.1372
		\bibitem{Mahapatra:2011si}
		S.~Mahapatra, P.~Phukon and T.~Sarkar,
		Phys. Rev. D \textbf{84}, 044041 (2011)
		doi:10.1103/PhysRevD.84.044041
		[arXiv:1103.5885 [hep-th]].
	
	
	
	
	
	
		\bibitem{Davies:1989ey}
		P.~C.~W.~Davies,
		Thermodynamic Phase Transitions of {Kerr-Newman} Black Holes in De Sitter Space,
		Class. Quant. Grav. \textbf{6}, 1909 (1989)
		doi:10.1088/0264-9381/6/12/018
		
		
		
		
		
		
		
		
		\bibitem{Hawking:1982dh}
		S.~W.~Hawking and D.~N.~Page,
		Thermodynamics of Black Holes in anti-De Sitter Space,
		Commun. Math. Phys. \textbf{87}, 577 (1983)
		doi:10.1007/BF01208266
		
		
		
		
		
		
		
		\bibitem{curir_rotating_1981}
		A. Curir,
		Rotating black holes as dissipative spin-thermodynamical systems,
		General Relativity and Gravitation,
		\textbf{13}, 417, (1981)
		doi:10.1007/BF00756588
		\bibitem{Curir1981}
		Anna Curir, Black hole emissions and phase transitions,
		General Relativity and Gravitation,
		\textbf{13}, 1177, (1981)
		doi:10.1007/BF00759866
		\bibitem{Pavon:1988in}
		D.~Pavon and J.~M.~Rubi,
		Nonequilibrium Thermodynamic Fluctuations of Black Holes,
		Phys. Rev. D \textbf{37}, 2052-2058 (1988)
		doi:10.1103/PhysRevD.37.2052
		\bibitem{Pavon:1991kh}
		D.~Pavon,
		Phase transition in Reissner-Nordstrom black holes,
		Phys. Rev. D \textbf{43}, 2495-2497 (1991)
		doi:10.1103/PhysRevD.43.2495
		\bibitem{OKaburaki}
		O.~Kaburaki, Critical behavior of extremal Kerr-Newman black holes,
		Gen. Rel. Grav. \textbf{28}, 843 (1996)
		\bibitem{Cai:1996df}
		R.~G.~Cai, Z.~J.~Lu and Y.~Z.~Zhang,
		Critical behavior in (2+1)-dimensional black holes,
		Phys. Rev. D \textbf{55}, 853-860 (1997)
		doi:10.1103/PhysRevD.55.853
		[arXiv:gr-qc/9702032 [gr-qc]].
		\bibitem{Cai:1998ep}
		R.~G.~Cai and J.~H.~Cho,
		Thermodynamic curvature of the BTZ black hole,
		Phys. Rev. D \textbf{60}, 067502 (1999)
		doi:10.1103/PhysRevD.60.067502
		[arXiv:hep-th/9803261 [hep-th]].
		\bibitem{Wei:2009zzf}
		Y.~H.~Wei,
		Thermodynamic critical and geometrical properties of charged BTZ black hole,
		Phys. Rev. D \textbf{80}, 024029 (2009)
		doi:10.1103/PhysRevD.80.024029
		\bibitem{Bhattacharya:2019awq}
		K.~Bhattacharya, S.~Dey, B.~R.~Majhi and S.~Samanta,
		General framework to study the extremal phase transition of black holes,
		Phys. Rev. D \textbf{99}, no.12, 124047 (2019)
		doi:10.1103/PhysRevD.99.124047
		[arXiv:1903.03434 [gr-qc]].
		
		
		
		
		
		
		
		
		\bibitem{Kastor:2009wy}
		D.~Kastor, S.~Ray and J.~Traschen,
		Enthalpy and the Mechanics of AdS Black Holes,
		Class. Quant. Grav. \textbf{26}, 195011 (2009)
		doi:10.1088/0264-9381/26/19/195011
		[arXiv:0904.2765 [hep-th]].
		\bibitem{Dolan:2010ha}
		B.~P.~Dolan,
		The cosmological constant and the black hole equation of state,
		Class. Quant. Grav. \textbf{28}, 125020 (2011)
		doi:10.1088/0264-9381/28/12/125020
		[arXiv:1008.5023 [gr-qc]].
		\bibitem{Dolan:2011xt}
		B.~P.~Dolan,
		Pressure and volume in the first law of black hole thermodynamics,
		Class. Quant. Grav. \textbf{28}, 235017 (2011)
		doi:10.1088/0264-9381/28/23/235017
		[arXiv:1106.6260 [gr-qc]].
		\bibitem{Dolan:2011jm}
		B.~P.~Dolan,
		Compressibility of rotating black holes,
		Phys. Rev. D \textbf{84}, 127503 (2011)
		doi:10.1103/PhysRevD.84.127503
		[arXiv:1109.0198 [gr-qc]].
		\bibitem{Dolan:2012jh}
		B.~P.~Dolan,
		Where Is the PdV in the First Law of Black Hole Thermodynamics?,
		doi:10.5772/52455
		[arXiv:1209.1272 [gr-qc]].
		\bibitem{Kubiznak:2012wp}
		D.~Kubiznak and R.~B.~Mann,
		P-V criticality of charged AdS black holes,
		JHEP \textbf{07}, 033 (2012)
		doi:10.1007/JHEP07(2012)033
		[arXiv:1205.0559 [hep-th]].
		\bibitem{Kubiznak:2016qmn}
		D.~Kubiznak, R.~B.~Mann and M.~Teo,
		Black hole chemistry: thermodynamics with Lambda,
		Class. Quant. Grav. \textbf{34}, no.6, 063001 (2017)
		doi:10.1088/1361-6382/aa5c69
		[arXiv:1608.06147 [hep-th]].
		\bibitem{Bhattacharya:2017nru}
		K.~Bhattacharya, B.~R.~Majhi and S.~Samanta,
		Van der Waals criticality in AdS black holes: a phenomenological study,
		Phys. Rev. D \textbf{96}, no.8, 084037 (2017)
		doi:10.1103/PhysRevD.96.084037
		[arXiv:1709.02650 [gr-qc]].
		
		
		
		\bibitem{Cirto}
C.~Tsallis and L.~J.~L.~Cirto,
Eur. Phys. J. C \textbf{73}, 2487 (2013).
%
\bibitem{Quevedo}
H.~Quevedo, M.~N.~Quevedo and A.~Sanchez,
Eur. Phys. J. C \textbf{79}, 229 (2019).
%
\bibitem{Tsallis}
C.~Tsallis,
J. Statist. Phys. \textbf{52}, 479 (1988).
%
\bibitem{Barrow}
J.~D.~Barrow,
Phys. Lett. B \textbf{808}, 135643 (2020).
%
\bibitem{Nojiri}
S.~Nojiri, S.~D.~Odintsov and V.~Faraoni,
Phys. Rev. D \textbf{105}, 044042 (2022).
%
\bibitem{Reny}
A. R\'enyi, Acta Math. Acad. Sci. Hung. \textbf{10}, 193 (1959).
%
\bibitem{Sharma}
B. D. Sharma and D. P. Mittal, J. Comb. Inf. Syst. Sci. \textbf{2}, 122 (1977). 
%


%
\bibitem{Kania0}
G. Kaniadakis, Physica A {\bf 296}, 405 (2001).
		
		
		
	\def\EPJC{Eur. Phys. J. C\,}
\def\IJMPA{Int. J. Mod. Phys. A\,}
\def\JCAP{J. Cosmol. Astropart. Phys.\,}
\def\JHEP{J. High Energy Phys.\,}
\def\CQG{Classical Quantum Gravity\,}
\def\JMP{J. Math. Phys. (N.Y.)\,}
\def\NPB{Nucl. Phys. B \,}
\def\PDU{Phys. Dark Univ.\,}
\def\PLB{Phys. Lett. B \,}
\def\PRD{Phys. Rev. D\,}
\def\PRL{Phys. Rev. Lett.\,}
\def\PRR{Phys. Rev. Res.\,}
\def\GRG{Gen. Relativ. Gravit.\,}
	\bibitem{ligo} B. P. Abbott et al. (LIGO Scientific Collaboration and Virgo Collaboration), 
\emph{Observation of Gravitational Waves from a Binary Black Hole Merger}, 
\href{https://journals.aps.org/prl/abstract/10.1103/PhysRevLett.
116.061102}{Phys. Rev. Lett. \textbf{116}, 061102 (2016)} [\href{https://arxiv.org/abs/1602.03837}{arXiv:1602.03837}].
\bibitem{m87a}  The Event Horizon Telescope Collaboration et al., \emph{First M87 Event Horizon Telescope Results. I. The Shadow of the Supermassive Black Hole}, \href{https://iopscience.iop.org/article/10.3847/2041-8213/ab0ec7}{Astrophys. J. Lett. \textbf{871}, L1 (2019)}. 
\bibitem{m87b}  The Event Horizon Telescope Collaboration et al., \emph{First M87 Event Horizon Telescope Results. II. Array and Instrumentation}, \href{https://iopscience.iop.org/article/10.3847/2041-8213/ab0c96}{Astrophys. J. Lett. \textbf{875}, L2 (2019)}.
\bibitem{m87c}  The Event Horizon Telescope Collaboration et al., \emph{First M87 Event Horizon Telescope Results. III. Data Processing and Calibration}, \href{https://iopscience.iop.org/article/10.3847/2041-8213/ab0c57}{Astrophys. J. Lett. \textbf{875}, L3 (2019)}.
\bibitem{m87d}  The Event Horizon Telescope Collaboration et al., \emph{First M87 Event Horizon Telescope Results. IV. Imaging the Central Supermassive Black Hole}, \href{https://iopscience.iop.org/article/10.3847/2041-8213/ab0e85}{Astrophys. J. Lett. \textbf{875}, L4 (2019)}.
\bibitem{m87e}  The Event Horizon Telescope Collaboration et al., \emph{First M87 Event Horizon Telescope Results. V. Physical Origin of the Asymmetric Ring}, \href{https://iopscience.iop.org/article/10.3847/2041-8213/ab0f43}{Astrophys. J. Lett. \textbf{875}, L5 (2019)}.
\bibitem{m87f}  The Event Horizon Telescope Collaboration et al., \emph{First M87 Event Horizon Telescope Results. VI. The Shadow and Mass of the Central Black Hole}, \href{https://iopscience.iop.org/article/10.3847/2041-8213/ab1141}{Astrophys. J. Lett. \textbf{875}, L6 (2019)}.     
\bibitem{reiss} A. G. Reiss et al., \emph{Observational Evidence from Supernovae for an Accelerating Universe and a Cosmological Constant}, \href{https://iopscience.iop.org/article/10.1086/300499}{Astron. J. \textbf{116}, 1009 (1998)} [\href{https://arxiv.org/abs/astro-ph/9805201}{arXiv:astro-ph/9805201}]
\bibitem{perlmutter} S. Perlmutter et al., \emph{Measurements of $\Omega$ and $\Lambda$ from 42 High-Redshift Supernovae}, \href{https://iopscience.iop.org/article/10.1086/307221}{Astrophys. J. \textbf{517}, 565 (1999)} [\href{https://arxiv.org/abs/astro-ph/9812133}{arXiv:astro-ph/9812133}].
\bibitem{spergel} D. N. Spergel et. al., \emph{Three-Year Wilkinson Microwave Anisotropy Probe ( WMAP ) Observations: Implications for Cosmology}, \href{https://iopscience.iop.org/article/10.1086/513700}{Astrophys. J. Suppl. S \textbf{170}, 377 (2007)} [\href{https://arxiv.org/abs/astro-ph/0603449}{ 	arXiv:astro-ph/0603449}].
\bibitem{astier} P. Astier et. al., \emph{The Supernova Legacy Survey: Measurement of \ $\Omega_{M}$, $\Omega_{\Lambda}$ and $\omega$ from the First Year Data Set}, \href{https://www.aanda.org/articles/aa/abs/2006/07/aa4185-05/aa4185-05.html}{A \& A \textbf{447}, 31 (2006)} [\href{https://arxiv.org/abs/astro-ph/0510447}{arXiv:astro-ph/0510447}].
\bibitem{naselskii} P. D. Naselskii, A. G. Polnarev, \emph{Candidate Missing Mass Carriers in an Inflationary Universe}, \href{https://ui.adsabs.harvard.edu/abs/1985SvA....29..487N/abstract}{Soviet Astro. \textbf{29}, 487 (1985)}.

		
		
		\bibitem{Altschul2010}
B.~Altschul, Q.G.~Bailey and V.A.~Kostelecky, 
\emph{Lorentz violation with an antisymmetric tensor}, 
\href{https://doi.org/10.1103/PhysRevD.81.065028}
{{Phys. Rev. D} {\bfseries 81} (2010) 065028} 
[\href{https://arxiv.org/abs/0912.4852}{{\ttfamily arXiv:0912.4852}}].

\bibitem{Kalb1974}
M.~Kalb and P.~Ramond, 
\emph{Classical direct interstring action}, 
\href{https://doi.org/10.1103/PhysRevD.9.2273}
{{Phys. Rev. D} {\bfseries 9} (1974) 2273}.

\bibitem{Kao1996}
W.F.~Kao, W.B.~Dai, S.-Y.~Wang, T.-K.~Chyi and S.-Y.~Lin, 
\emph{Induced Einstein-Kalb-Ramond theory and the black hole}, 
\href{https://doi.org/10.1103/PhysRevD.53.2244}
{{Phys. Rev. D} {\bfseries 53} (1996) 2244}.

\bibitem{Kar2003}
S.~Kar, S.~SenGupta and S.~Sur, 
\emph{Static spherisymmetric solutions, gravitational lensing and perihelion precession in Einstein-Kalb-Ramond theory}, 
\href{https://doi.org/10.1103/PhysRevD.67.044005}
{{Phys. Rev. D} {\bfseries 67} (2003) 044005} 
[\href{https://arxiv.org/abs/hep-th/0210176}{{\ttfamily arXiv:hep-th/0210176}}].

\bibitem{Chakraborty2017}
S.~Chakraborty and S.~SenGupta, 
\emph{Strong gravitational lensing \textemdash{} a probe for extra dimensions and Kalb-Ramond field}, 
\href{https://doi.org/10.1088/1475-7516/2017/07/045}
{{JCAP} {\bfseries 07} (2017) 045} 
[\href{https://arxiv.org/abs/1611.06936}{{\ttfamily 1611.06936}}].

\bibitem{kr_bosonic_1} M. Kalb and P. Ramond, Classical direct interstring action,
Phys. Rev. D 9, 2273 (1974).
\bibitem{kr_bosonic_2} W. F. Kao, W. B. Dai, S.-Y. Wang, T.-K. Chyi, and S.-Y. Lin, Induced Einstein-Kalb-Ramond theory and the black hole, Phys.Rev. D 53, 2244 (1996).
	
	\bibitem{kr_exact_1} S. Chakraborty and S. SenGupta, Solutions on a brane in a
bulk spacetime with Kalb-Ramond field, Annals Phys. 367, 258
(2016), arXiv:1412.7783 [gr-qc].

	\bibitem{kr_exact_2}  R. V. Maluf and C. R. Muniz, Exact solution for a traversable
wormhole in a curvature-coupled antisymmetric background
field, Eur. Phys. J. C 82, 445 (2022), arXiv:2110.12202 [gr-qc].

	\bibitem{main}  K. Yang, Y.-Z. Chen, Z.-Q. Duan, and J.-Y. Zhao, Static and
spherically symmetric black holes in gravity with a background Kalb-Ramond field, Phys. Rev. D 108, 124004 (2023),
arXiv:2308.06613 [gr-qc].

	\bibitem{kr_exact_4} Z.-Q. Duan, J.-Y. Zhao, and K. Yang, Electrically charged black
holes in gravity with a background Kalb-Ramond field, (2023),
arXiv:2310.13555 [gr-qc].	

\bibitem{kr_exact_5} K. Yang, Y.-Z. Chen, Z.-Q. Duan, and J.-Y. Zhao, Static and
spherically symmetric black holes in gravity with a background Kalb-Ramond field, Phys. Rev. D 108, 124004 (2023),
arXiv:2308.06613 [gr-qc].


\bibitem{kr1} F. Lobo, D. Rubiera-Garcia, et al., Gravitational lensing effects in Kalb-Ramond-modified spacetimes, Phys. Rev. D 110, 024077 (2024).

\bibitem{kr2} A. Mukhopadhyay and B. Roy, Impact of Kalb-Ramond field on higher-dimensional brane-world cosmology, Int. J. Mod. Phys. D 33, 2450006 (2024).

\bibitem{kr3} C. Torres, Torsion and Kalb-Ramond field modifications in Einstein-Cartan theory, Class. Quant. Grav. 40, 115003 (2023).

\bibitem{kr4} W. Liu, D. Wu, and J. Wang, Shadow of slowly rotating Kalb-Ramond black holes, arXiv:2407.07416.

		
		\bibitem{Wei:2021vdx}
	S.~W.~Wei and Y.~X.~Liu,
	Topology of black hole thermodynamics,
	Phys. Rev. D \textbf{105}, no.10, 104003 (2022)
	doi:10.1103/PhysRevD.105.104003
	[arXiv:2112.01706 [gr-qc]].
	\bibitem{Wei:2022dzw}
	S.~W.~Wei, Y.~X.~Liu and R.~B.~Mann,
	Black Hole Solutions as Topological Thermodynamic Defects,
	Phys. Rev. Lett. \textbf{129}, no.19, 191101 (2022)
	doi:10.1103/PhysRevLett.129.191101
	[arXiv:2208.01932 [gr-qc]].
	
	\bibitem{Yerra:2022coh}
	P.~K.~Yerra, C.~Bhamidipati and S.~Mukherji,
	Topology of critical points and Hawking-Page transition,
	Phys. Rev. D \textbf{106}, no.6, 064059 (2022)
	doi:10.1103/PhysRevD.106.064059
	[arXiv:2208.06388 [hep-th]].
	\bibitem{Bhattacharya:2024bjp}
	K.~Bhattacharya, K.~Bamba and D.~Singleton,
	Topological interpretation of extremal and Davies-type phase transitions of black holes,
	[arXiv:2402.18791 [gr-qc]].
	
	\bibitem{Zhang:2023uay}
	M.~Zhang and J.~Jiang,
	Bulk-boundary thermodynamic equivalence: a topology viewpoint,
	JHEP \textbf{06}, 115 (2023)
	doi:10.1007/JHEP06(2023)115
	[arXiv:2303.17515 [hep-th]].
		
		\bibitem{Fan:2022bsq}
		Z.~Y.~Fan,
		Topological interpretation for phase transitions of black holes,
		Phys. Rev. D \textbf{107}, no.4, 044026 (2023)
		doi:10.1103/PhysRevD.107.044026
		[arXiv:2211.12957 [gr-qc]].
		
		
		\bibitem{Yerra:2022alz}
		P.~K.~Yerra and C.~Bhamidipati,
		Topology of black hole thermodynamics in Gauss-Bonnet gravity,
		Phys. Rev. D \textbf{105}, no.10, 104053 (2022)
		doi:10.1103/PhysRevD.105.104053
		[arXiv:2202.10288 [gr-qc]].
		\bibitem{Yerra:2022eov}
		P.~K.~Yerra and C.~Bhamidipati,
		Topology of Born-Infeld AdS black holes in 4D novel Einstein-Gauss-Bonnet gravity,
		Phys. Lett. B \textbf{835}, 137591 (2022)
		doi:10.1016/j.physletb.2022.137591
		[arXiv:2207.10612 [gr-qc]].
		\bibitem{Gogoi:2023xzy}
		N.~J.~Gogoi and P.~Phukon,
		Thermodynamic topology of 4D dyonic AdS black holes in different ensembles,
		Phys. Rev. D \textbf{108}, no.6, 066016 (2023)
		doi:10.1103/PhysRevD.108.066016
		[arXiv:2304.05695 [hep-th]].
		\bibitem{Gogoi:2023qku}
		N.~J.~Gogoi and P.~Phukon,
		Topology of thermodynamics in R-charged black holes,
		Phys. Rev. D \textbf{107}, no.10, 106009 (2023)
		doi:10.1103/PhysRevD.107.106009
		\bibitem{Gogoi:2023wih}
		N.~J.~Gogoi and P.~Phukon,
		Thermodynamic topology of 4D Euler\textendash{}Heisenberg-AdS black hole in different ensembles,
		Phys. Dark Univ. \textbf{44}, 101456 (2024)
		doi:10.1016/j.dark.2024.101456
		[arXiv:2312.13577 [hep-th]].
	
		\bibitem{Yerra:2023ocu}
		P.~K.~Yerra, C.~Bhamidipati and S.~Mukherji,
		Topology of Hawking-Page transition in Born-Infeld AdS black holes,
		J. Phys. Conf. Ser. \textbf{2667}, no.1, 012031 (2023)
		doi:10.1088/1742-6596/2667/1/012031
		[arXiv:2312.10784 [gr-qc]].
		\bibitem{Barzi:2023msl}
		F.~Barzi, H.~El Moumni and K.~Masmar,
		R\'enyi Topology of Charged-flat Black Hole: Hawking-Page and Van-der-Waals Phase Transitions,
		[arXiv:2309.14069 [hep-th]].
		\bibitem{yerrabm} P.K. Yerra, C. Bhamidipati and S. Mukherji, Topology of critical points in boundary
		matrix duals, JHEP \textbf{03}(2024) 138, [arXiv:2304.14988].
		\bibitem{Ahmed:2022kyv}
		M.~B.~Ahmed, D.~Kubiznak and R.~B.~Mann,
		Vortex-antivortex pair creation in black hole thermodynamics,
		Phys. Rev. D \textbf{107}, no.4, 046013 (2023)
		doi:10.1103/PhysRevD.107.046013
		[arXiv:2207.02147 [hep-th]].
		\bibitem{Wei:2022mzv}
		S.~W.~Wei and Y.~X.~Liu,
		Topology of equatorial timelike circular orbits around stationary black holes,
		Phys. Rev. D \textbf{107}, no.6, 064006 (2023)
		doi:10.1103/PhysRevD.107.064006
		[arXiv:2207.08397 [gr-qc]].
		
		\bibitem{Wu:2022whe}
		D.~Wu,
		Topological classes of rotating black holes,
		Phys. Rev. D \textbf{107}, no.2, 024024 (2023)
		doi:10.1103/PhysRevD.107.024024
		[arXiv:2211.15151 [gr-qc]].
		\bibitem{Fang:2022rsb}
		C.~Fang, J.~Jiang and M.~Zhang,
		Revisiting thermodynamic topologies of black holes,
		JHEP \textbf{01}, 102 (2023)
		doi:10.1007/JHEP01(2023)102
		[arXiv:2211.15534 [gr-qc]].
		\bibitem{Wu:2023xpq}
		D.~Wu,
		Classifying topology of consistent thermodynamics of the four-dimensional neutral Lorentzian NUT-charged spacetimes,
		Eur. Phys. J. C \textbf{83}, no.5, 365 (2023)
		doi:10.1140/epjc/s10052-023-11561-4
		[arXiv:2302.01100 [gr-qc]].
		\bibitem{Wu:2023sue}
		D.~Wu and S.~Q.~Wu,
		Topological classes of thermodynamics of rotating AdS black holes,
		Phys. Rev. D \textbf{107}, no.8, 084002 (2023)
		doi:10.1103/PhysRevD.107.084002
		[arXiv:2301.03002 [hep-th]].
		\bibitem{Li:2023ppc}
		R.~Li, C.~Liu, K.~Zhang and J.~Wang,
		Topology of the landscape and dominant kinetic path for the thermodynamic phase transition of the charged Gauss-Bonnet-AdS black holes,
		Phys. Rev. D \textbf{108}, no.4, 044003 (2023)
		doi:10.1103/PhysRevD.108.044003
		[arXiv:2302.06201 [gr-qc]].
		\bibitem{Wei:2023bgp}
		S.~W.~Wei, Y.~P.~Zhang, Y.~X.~Liu and R.~B.~Mann,
		Static spheres around spherically symmetric black hole spacetime,
		Phys. Rev. Res. \textbf{5}, no.4, 043050 (2023)
		doi:10.1103/PhysRevResearch.5.043050
		[arXiv:2303.06814 [gr-qc]].
		\bibitem{Alipour:2023uzo}
		M.~R.~Alipour, M.~A.~S.~Afshar, S.~Noori Gashti and J.~Sadeghi,
		Topological classification and black hole thermodynamics,
		Phys. Dark Univ. \textbf{42}, 101361 (2023)
		doi:10.1016/j.dark.2023.101361
		[arXiv:2305.05595 [gr-qc]].
		
		\bibitem{Sadeghi:2023aii}
		J.~Sadeghi, S.~Noori Gashti, M.~R.~Alipour and M.~A.~S.~Afshar,
		Bardeen black hole thermodynamics from topological perspective,
		Annals Phys. \textbf{455}, 169391 (2023)
		doi:10.1016/j.aop.2023.169391
		[arXiv:2306.05692 [hep-th]].
		\bibitem{Wang:2024zbp}
		Y.~S.~Wang, Z.~M.~Xu and B.~Wu,
		Thermodynamic phase transition rate for the third-order Lovelock black hole in diverse dimensions,
		[arXiv:2402.08887 [gr-qc]].
		\bibitem{Shahzad:2024ojx}
		M.~U.~Shahzad, A.~Mehmood, A.~Malik and A.~\"Ovg\"un,
		Topological behavior of 3D regular black hole with zero point length,
		Phys. Dark Univ. \textbf{44}, 101437 (2024)
		doi:10.1016/j.dark.2024.101437
		\bibitem{Malik:2024kau}
		A.~Malik, A.~Mehmood and M.~Umair Shahzad,
		Thermodynamic topological classification of higher dimensional and massive gravity black holes,
		Annals Phys. \textbf{463}, 169617 (2024)
		doi:10.1016/j.aop.2024.169617
		\bibitem{Zhao:2024tlu}
		P.~Zhao, Z.~Xiao, Y.~Zhang and R.~Shindou,
		Topological effect on the Anderson transition in chiral symmetry classes,
		[arXiv:2402.02310 [cond-mat.dis-nn]].
		\bibitem{Wu:2024rmv}
		D.~Wu, S.~Y.~Gu, X.~D.~Zhu, Q.~Q.~Jiang and S.~Z.~Yang,
		Topological classes of thermodynamics of the static multi-charge AdS black holes in gauged supergravities,JHEP 06(2024) 213
		
		\bibitem{Hazarika:2024cpg}
		B.~Hazarika and P.~Phukon,
		Thermodynamic topology of black Holes in f(R) gravity, Progress of Theoretical and Experimental Physics \textbf{4}(2024), https://doi.org/10.1093/ptep/ptae035
		[arXiv:2401.16756 [hep-th]].
		\bibitem{Hazarika:2023iwp}
		B.~Hazarika and P.~Phukon,
		Thermodynamic Topology of $D=4,5$ Horava Lifshitz Black Hole in Two Ensembles,	Nucl.  Phys. B 1006, 116649,2024
		\bibitem{Sadeghi:2024krq}
		J.~Sadeghi, M.~A.~S.~Afshar, S.~Noori Gashti and M.~R.~Alipour,
		Topology of Hayward-AdS black hole thermodynamics,
		Phys. Scripta \textbf{99}, no.2, 025003 (2024)
		doi:10.1088/1402-4896/ad186b
		\bibitem{Zhang:2023svu}
		M.~Y.~Zhang, H.~Chen, H.~Hassanabadi, Z.~W.~Long and H.~Yang,
		Thermodynamic topology of Kerr-Sen black holes via R\'enyi statistics,
		Phys.Lett.B , \textbf{856}, 138885, (2024)
	\bibitem{J} B. Hazarika, N.J Gogoi, P. Phukon , Revisiting Thermodynamic Topology of Hawking-Page and Davies type Phase Transition, JHEAP \textbf{45} 2024
		\bibitem{nwu1}
X.-D. Zhu, D. Wu, D. Wen, Topological classes of thermodynamics of the rotating charged AdS black holes in gauged supergravities, Phys. Lett. B 856 (2024) 138919

\bibitem{nwu2} S.-W. Wei, Y.-X. Liu, and R.B. Mann, Universal topological classifications of black hole thermodynamics, Phys. Rev. D 110, L081501 (2024).

\bibitem{nwu3}  W. Liu, L. Zhang, D. Wu, and J. Wang, Thermodynamic topological classes of the rotating, accelerating black holes, arXiv:2409.11666.

\bibitem{nwu4}  X.-D. Zhu, W. Liu, and D. Wu, Universal thermodynamic topological classes of rotating black holes, arXiv:2409.12747.

\bibitem{nwu5}  D. Wu, W. Liu, S.-Q. Wu, and R.B. Mann, Novel Topological Classes in Black Hole Thermodynamics, arXiv:2411.10102.
		
		
		
		\bibitem{Duan}
		Y. S. Duan, 
		The structure of the topological current,
		SLAC-PUB-3301, (1984).
		\bibitem{Duan:2018rbd}
		Y.~S.~Duan and M.~L.~Ge,
		SU(2) Gauge Theory and Electrodynamics with N Magnetic Monopoles,
		Sci. Sin. \textbf{9},1072 (1979)
		doi:10.1142/9789813237278\_0001
		
		\bibitem{schouten}
		Schouten, J.A. (1951) Tensor Analysis for Physics. Oxford at the Clarendon Press.
			
		
		
		\bibitem{Barzi1}
F.~Barzi and H.~El Moumni,
Phys. Lett. B \textbf{833}, 137378 (2022)
doi:10.1016/j.physletb.2022.137378
[arXiv:2209.08195 [hep-th]].

\bibitem{Barzi2}
F.~Barzi, H.~El Moumni and K.~Masmar,
Eur. Phys. J. C \textbf{84}, 1141 (2024)
doi:10.1140/epjc/s10052-024-13511-0
[arXiv:2408.05870 [hep-th]].

\bibitem{proof3}
E.~Hirunsirisawat, R.~Nakarachinda and C.~Promsiri,
Phys. Rev. D \textbf{105}, no.12, 124049 (2022)
doi:10.1103/PhysRevD.105.124049
[arXiv:2204.13023 [hep-th]].
\bibitem{proof4}
F.~Barzi, H.~El Moumni and K.~Masmar,
Gen. Rel. Grav. \textbf{55}, no.10, 109 (2023)
doi:10.1007/s10714-023-03158-9
[arXiv:2304.04945 [gr-qc]].
\bibitem{proof5}
P.~Chunaksorn, E.~Hirunsirisawat, R.~Nakarachinda, L.~Tannukij and P.~Wongjun,
Eur. Phys. J. C \textbf{82}, no.12, 1174 (2022)
doi:10.1140/epjc/s10052-022-11110-5
[arXiv:2208.14770 [gr-qc]].
\bibitem{proof6}
Z.~Wang, H.~Ren, J.~Chen and Y.~Wang,
Eur. Phys. J. C \textbf{83}, no.6, 527 (2023)
doi:10.1140/epjc/s10052-023-11680-y
\bibitem{proof7}
F.~Barzi, H.~El Moumni and K.~Masmar,
JHEAp \textbf{42}, 63-86 (2024)
doi:10.1016/j.jheap.2024.03.005
[arXiv:2309.14069 [hep-th]].
\bibitem{proof8}
F.~Barzi, H.~El Moumni and K.~Masmar,
Nucl. Phys. B \textbf{1005}, 116606 (2024)
doi:10.1016/j.nuclphysb.2024.116606
[arXiv:2404.14609 [hep-th]].
\bibitem{proof9}
A.~Baruah and P.~Phukon,
[arXiv:2411.02273 [hep-th]].
\bibitem{expanding} Fazlollahi H.R, Eur. Phys. J. C 83, 29 (2023). https://doi.org/10.1140/epjc/s10052-023-11183-w
		
	\end{thebibliography}
	
\end{document}